\documentclass[journal]{IEEEtran}
\IEEEoverridecommandlockouts
\usepackage{textcomp}
\usepackage{stfloats}
\usepackage[utf8]{inputenc}
\usepackage{balance}
\usepackage{caption,subcaption}
\usepackage{float}

\usepackage{amsmath,amssymb,amsfonts,bm}
\usepackage{algorithmic}
\usepackage[ruled,vlined]{algorithm2e}

\usepackage{pgfplots}
\usepackage{grffile}
\pgfplotsset{compat=newest}
\usetikzlibrary{math,calc,plotmarks,arrows.meta}
\usepackage{graphicx}
\usepackage{xcolor} % More colours

\usepackage{cite}
\usepackage{hyperref}
\usepackage{siunitx}

\def\BibTeX{{\rm B\kern-.05em{\sc i\kern-.025em b}\kern-.08em
    T\kern-.1667em\lower.7ex\hbox{E}\kern-.125emX}}

\DeclareMathOperator*{\argmax}{arg\,max}
\DeclareMathOperator*{\argmin}{arg\,max}
\DeclareMathOperator*{\R}{\mathbb{R}}
\DeclareMathOperator*{\E}{\mathbb{E}}
\DeclareMathOperator*{\p}{\mathbb{P}}
\DeclareMathOperator{\atanh}{atanh}
\renewcommand{\Vec}{\mathbf}

\begin{document}
\title{Post-processing of EEG-based Auditory Attention Decoding Decisions via Hidden Markov Models\\
\thanks{N. Heintz and A. Bertrand are with KU Leuven, Department of Electrical Engineering (ESAT), STADIUS Center for Dynamical Systems, Signal Processing and Data Analytics and also with Leuven.AI - KU Leuven institute for AI, Kasteelpark Arenberg 10, B-3001 Leuven, Belgium (e-mail: nicolas.heintz@esat.kuleuven.be, alexander.bertrand@esat.kuleuven.be). 

N. Heintz  and T. Francart are with KU Leuven, Department of Neurosciences, Research Group ExpORL, Herestraat 49 box 721, B-3000 Leuven, Belgium (e-mail: tom.francart@kuleuven.be).

The authors acknowledge the financial support of the FWO (Research Foundation Flanders) for project G081722N and 1S31524N, the Flemish Government (AI Research Program), Internal Funds KU Leuven (project IDN/23/006 and C14/25/108), and the European Research Council (ERC) under the European Union’s research and innovation programme (grant agreement No 101138304). Views and opinions expressed are however those of the author(s) only and do not necessarily reflect those of the European Union or any of the granting authorities. Neither the European Union nor the granting authorities can be held responsible for them.
}
}

% Authors
\author{Nicolas~Heintz, Tom~Francart, and~Alexander~Bertrand}

\maketitle
% Abstract
\begin{abstract}
Auditory attention decoding (AAD) algorithms exploit brain signals, such as electroencephalography (EEG), to identify which speaker a listener is focusing on in a multi-speaker environment. While state-of-the-art AAD algorithms can identify the attended speaker on short time windows, their predictions are often too inaccurate for practical use. In this work, we propose augmenting AAD with a hidden Markov model (HMM) that models the temporal structure of attention. 
More specifically, the HMM relies on the fact that a subject is much less likely to switch attention than to keep attending the same speaker at any moment in time. We show how a HMM can significantly improve existing AAD algorithms in both causal (real-time) and non-causal (offline) settings. We further demonstrate that HMMs outperform existing postprocessing approaches in both accuracy and responsiveness, and explore how various factors such as window length, switching frequency, and AAD accuracy influence overall performance. The proposed method is computationally efficient, intuitive to use and applicable in both real-time and offline settings.
\end{abstract}


%% Introduction %%
\section{Introduction}
\label{sec:Introduction}
Accurately detecting to whom someone wishes to listen is of crucial importance for a wide array of applications. For example, this would allow a hearing aid to determine which speakers should be enhanced or suppressed \cite{Geirnaert2020a, nguyen_aadnet_2024, straetmans_neural_2024, wilroth_improving_2025}. This problem can potentially be solved by decoding the auditory attention from brain signals using electroencephalography (EEG) \cite{OSullivan2014, Biesmans2017, DeCheveigne2018, Miran2018, Heintz2023Unbiased}. 

The most common and reliable method to decode attention from the neural response is based on stimulus reconstruction \cite{OSullivan2014,Biesmans2017,DeCheveigne2018,straetmans_neural_2024,thornton_robust_2022}. This method is based on the observation that the brain tracks attended speech more than unattended speech \cite{Aiken2008,Ding2012}. The goal is to train a decoder that reconstructs the temporal variations in the attended speech signal (e.g., its amplitude envelope) from the EEG data. It is thus possible to reconstruct attended speech from the neural response. The speech signal that is most similar to this reconstruction is then predicted to be the attended speaker. Other AAD algorithms have since emerged that decode the location of the attended speaker \cite{Geirnaert2020CSP, Vandecapelle2020}, or classify the auditory attention directly using neural networks \cite{nguyen_aadnet_2024}. However, it was recently shown that such models tend to heavily overfit on trial fingerprints, eye gaze, or other dataset-dependent shortcuts, making them unreliable in uncontrolled environments \cite{rotaru_what_2024, Puffay2023}.

AAD algorithms typically cut the EEG signal into smaller pieces and then classify the attended speaker per window. Ideally, the algorithm should achieve high accuracy within a short window to achieve a high temporal resolution in the AAD decisions. Using short windows ensures that a sudden switch in attention is promptly detected. This is nevertheless challenging, as most AAD algorithms -and especially the more reliable stimulus reconstruction algorithms- barely perform better than chance on very short window lengths. 

One potential solution is to smooth these inaccurate predictions using postprocessing techniques. While such smoothing techniques inevitably reduce the time resolution of AAD predictions, it was shown that the predictions of such postprocessing algorithms can be more accurate and faster than simply predicting the attended speaker using longer windows \cite{Geirnaert2020, heintz_probabilistic_2024}. Such techniques have therefore recently gained popularity to create real-time AAD systems, where smooth, yet responsive predictions are vital \cite{Miran2018,aroudi_closed-loop_2021,heintz_probabilistic_2024}. These algorithms mostly target real-time applications, where only past EEG data is available. While it is possible to use similar algorithms for offline applications, where both past and future data are available, existing AAD postprocessing algorithms are not capable of exploiting the access to past and future data to the fullest extent. 

Neighbouring data is valuable because the attention process of a listener is far from random. While the listener could switch attention on \textit{any} second, creating the necessity of short window lengths, it is highly unlikely that a listener will switch attention after \textit{every} second. Instead, in a natural scenario, a listener is expected to attend to the same speaker or conversation for a few minutes. The core idea explored in this paper is to model this attention process mathematically using a hidden Markov model (HMM) \cite{baum_maximization_1970}. Such a model enables us to compute the probability that any predicted attention trajectory corresponds to the listener's true attention trajectory. Hidden Markov models will typically favour simple trajectories with relatively few attention switches over more complex trajectories where the attended speaker switches every few seconds. 

Hidden Markov models have two main benefits. The first is their low computational cost. Even seemingly difficult tasks, such as computing the average probability that a listener was listening to a specific speaker at a specific time, given all available data, or computing the most likely attention trajectory, can be solved in linear time. The second benefit is their versatility. The model can be added to any existing AAD algorithm, as long as its predictions are approximately independent from window to window when conditioned on the attention state. It is also intuitive to adapt the model to a specific listening scenario, e.g., to add additional competing speakers or to add more complex interactions, while keeping the number of hyperparameters low. 

In Sections \ref{sec:HMM} and \ref{sec:AAD}, we will further explain how a hidden Markov model can be applied to an AAD task. Section \ref{sec:Experiments} contains the exact details of all performed experiments. The results of these experiments are reported in Section \ref{sec:results}. Throughout the paper, we will always differentiate between causal and non-causal hidden Markov models. Causal models only have access to past data, while non-causal models have access to both past and future data. We will show that a hidden Markov model-based AAD algorithm can both accurately detect the attended speaker and switch quickly when the listener switches attention. This is even true when the model has only access to past data (making it suited for real-time applications).

%% Hidden Markov model theory
\section{Hidden Markov Model}
\label{sec:HMM}
Our proposed hidden Markov model-based (HMM) AAD algorithm is a post-processing algorithm that uses the predictions of any AAD algorithm as input, i.e., it is agnostic to the specific AAD algorithm with which it is combined. This AAD algorithm is assumed to split the EEG into windows and predicts, per window, who the attended speaker is based on the EEG data in that window. In Section \ref{sec:AAD}, we will discuss two such AAD algorithms that will be used in the experiments in this paper. The HMM module then computes the probability that speaker $s$ is attended in a specific window $t$ based on all previous (and optionally future) AAD predictions. This enables us to infer who the attended speaker is in a specific window based on all available data, instead of only the data in a specific window. For example, if the AAD algorithm is confident that the listener attends to speaker $s$ in all windows surrounding a specific window $t$, the HMM will increase the probability that the listener is listening to that speaker in window $t$.

In real-time scenarios, where only previous data is available, this creates a smoothing effect on decisions at the cost of a slight delay, similar to the gain control algorithms proposed in \cite{Geirnaert2020,Aroudi2020, heintz_probabilistic_2024}. However, the HMM model truly shines in offline applications, where it has access to both previous and future data, e.g., for applications or experiments where a post-hoc analysis of the attention trajectory is relevant. We will show that the HMM can leverage this data to detect switches with a high temporal resolution and substantially fewer decision errors than the underlying AAD algorithm.

In this section, we will first explain how an AAD recording can be represented in a Hidden Markov Model. We will then explore how such a representation of the data can be used to improve the performance of AAD algorithms in both real-time (causal) and offline (non-causal) situations.

\subsection{Basic structure of a Hidden Markov Model for AAD}
AAD algorithms estimate a $D$-dimensional feature or score $\Vec{x}(t)\in \R^D$ that indicates to which speaker a listener attends in window $t$. $\Vec{x}(t)$ can be a hard decision (e.g. a single binary variable in the case of a 2-speaker setting), a soft decision (e.g. a probability vector for each speaker), or a feature vector from which the attended speaker can be inferred (e.g. the correlation between the decoded neural signals and each speech signal). In Section \ref{sec:AAD} we will provide a few examples. At this point, it is sufficient to treat these scores as observations that depend on an underlying attention process; we expect the statistics of $\Vec{x}(t)$ to change depending on who the attended speaker is. 

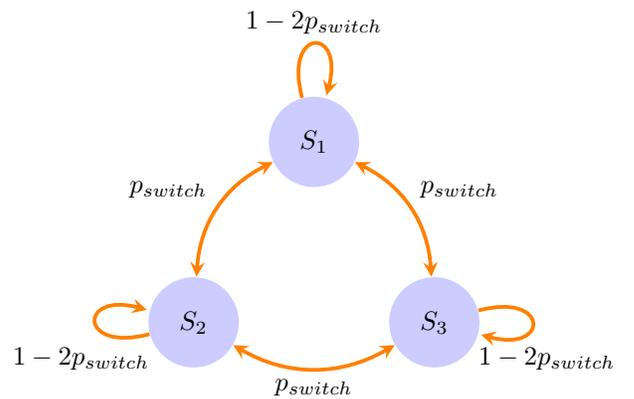
\begin{figure}[ht]
\centering   
\begin{tikzpicture}[scale=0.8,node distance=2.5cm, >=stealth, auto]
    % Define states in an equilateral triangle
    \node[circle, draw=none, fill=blue!20, minimum size=1.2cm] (S1) at (0, 2) {$S_1$};
    \node[circle, draw=none, fill=blue!20,minimum size=1.2cm] (S2) at (-2, -1) {$S_2$};
    \node[circle, draw=none, fill=blue!20,minimum size=1.2cm] (S3) at (2, -1) {$S_3$};

   % Transitions arrows
    \path[<->, line width = 0.5mm] (S1) edge[bend right, draw=orange] node[above left] {$p_{switch}$} (S2);
    \path[<->, line width = 0.5mm] (S2) edge[bend right, draw=orange] node[below] {$p_{switch}$} (S3);
    \path[<->, line width = 0.5mm] (S3) edge[bend right, draw=orange] node[above right] {$p_{switch}$} (S1);

    \path[->, line width = 0.5mm] (S1) edge[loop above, draw=orange] node[above] {$1 - 2p_{switch}$} (S1);
    \path[->, line width = 0.5mm] (S2) edge[loop left, draw=orange] node[below, xshift = -2mm, yshift= -2mm] {$1 - 2p_{switch}$} (S2);
    \path[->, line width = 0.5mm] (S3) edge[loop right, draw=orange] node[below, xshift = 2mm, yshift=-2mm] {$1 - 2p_{switch}$} (S3);
\end{tikzpicture}
\caption{The proposed Markov chain for three speakers and equal transition probabilities $p_{switch}$. Typically, $p_{switch} \ll 1/N$ in a scenario with $N$ speakers.}
\label{fig:HMM_model}
\end{figure}

This attention process is highly structured in reality: listeners are much more likely to listen to the same speaker for a long time than to frequently switch between speakers. Such prior knowledge can be modelled as a Markov chain, where each speaker is a node, and the edge between two nodes represents the probability for a listener to switch between the two speakers. In scenarios where there are multiple simultaneous conversations, we treat each conversation as a single speaker and combine the speech of all conversation participants into a single speech \cite{van_de_ryck_eeg-based_2025}. As such, switches between speakers within the same conversation are not penalised. An example of a Markov chain for three speakers is shown in Figure \ref{fig:HMM_model}. The probability of observing a certain series of attended speakers $\Vec{S}_{0:T} = [S(0) \dots S(T)]$ in windows $0$ to $T$ is then: 
\begin{equation}
\p(\Vec{S}_{0:T}) = \p(S(0))\prod_{t=1}^T \p(S(t)|S(t-1)), 
\label{eq:ProbKnownState}
\end{equation}
with $S(t)$ the attended speaker in window $t$.

For this paper, we chose a relatively simple Markov chain where each speaker is a priori equally likely to be attended and the probability of switching between speakers is equal for all speakers. The switching probability is a hyperparameter that is typically chosen such that a listener is much more likely to keep listening to the same speaker than to switch between speakers. 

A Hidden Markov Model combines the imposed structure from the Markov chain with the observations $\Vec{x}(t)$ from the AAD model to compute the probability of observing $\Vec{x}_{0:T} = [\Vec{x}(0)\dots\Vec{x}(T)]$ given the underlying attention process $\Vec{S}_{0:T} = [S(0) \dots S(T)]$. This is done by first modelling the probability $\p(\Vec{x}(t) | S(t))$ that observation $\Vec{x}(t)$ is observed in state $S(t)$ for each state $S_j,\ j\in [1,N]$ with $N$ the total number of speakers. The probability to observe $\Vec{x}$ given the attention process $\Vec{S}$ is then: 
\begin{align}
    \p(\Vec{x}_{0:T}) &= \p(\Vec{x}_{0:T}|\Vec{S}_{0:T})\p(\Vec{S}_{0:T})\\
                &= \p(\Vec{S}_{0:T})\prod_{t=0}^T \p(\Vec{x}(t)|S(t)),
    \label{eq:ActualUnknownHMMProb}
\end{align}
where we assume that the observations $\Vec{x}(t)$ only depend on the state of attention in window $t$. Note that \eqref{eq:ActualUnknownHMMProb} also assumes that $\Vec{x}(t)$ is independent from other observations $\Vec{x}(t-j)$ for $j\in \mathbb{Z} \setminus \{0\}$ when conditioned on $S(t)$, i.e., $\p(\Vec{x}(t)|S(t),\Vec{x}_{0:t-1}) = \p(\Vec{x}(t)|S(t))$.

The exact attention trajectory $\Vec{S}_{0:T}$ is naturally unknown in AAD, as inferring this trajectory is the main goal of AAD. Instead, we compute $\p(\hat{\Vec{S}}_{0:T}|\Vec{x}_{0:T})$, the probability that a given estimated attention trajectory $\hat{\Vec{S}}_{0:T}$ corresponds to the actual attention trajectory $\Vec{S}_{0:T}$ of the underlying HMM given the observed AAD scores $\Vec{x}_{0:T}$. We can then compute the probability of each possible path $\hat{\Vec{S}}_{0:T}$ and compute the probability-weighted average of all possible paths to obtain the probability $\p(\hat{S}(t)=S_j|\Vec{x}_{0:T}) \forall t$ that a listener attended to speaker $S_j$ in window $t$. Although an exhaustive search over all possible paths would be computationally infeasible, Sections \ref{subsec:causalInference}-\ref{subsec:Viterbi} will demonstrate how this process can be done efficiently. We consider three distinct cases: In Section \ref{subsec:causalInference}, we assume that only observations before window $t$ are available and thus compute $\p(\hat{S}(t)=S_j|\Vec{x}_{0:t})$. This is mostly relevant for real-time scenarios, where future observations haven't been measured yet. In Section \ref{subsec:non-causalInference}, we compute $\p(\hat{S}(t)=S_j|\Vec{x}_{0:T})$, which is more relevant for offline analysis where future observations are also available. Finally, we explain in Section \ref{subsec:Viterbi} that it is surprisingly cheap to find the entire attention process $\hat{\Vec{S}}$ that maximises $\p(\hat{\Vec{S}}|\Vec{x})$ in a HMM using the Viterbi algorithm \cite{viterbi_error_1967}. This is mostly useful when one is only interested in a single discrete prediction of the attended speaker per window, rather than a set of probabilities. 

\subsection{Causal inference of the attended speaker}
\label{subsec:causalInference}
The conditional probability $\p(\hat{S}(t)=S_j|[\Vec{x}(0)\dots\Vec{x}(t)])$ that the listener attends to speaker $S_j$ in window $t$ can be computed with the forward algorithm \cite{baum_maximization_1970}. First, we will compute the joint probability $\p(\hat{S}(t)=S_j,[\Vec{x}(0)\dots\Vec{x}(t)])$. For notational convenience, we define
\begin{align}
    \alpha_j(t) &\triangleq \p\left(\hat{S}(t) = S_j,\Vec{x}_{0:t}\right) \\
    a_{jk}      &\triangleq \p\left(\hat{S}(t) = S_j| \hat{S}(t-1) = S_k\right)\\
    b_{j}(t)    &\triangleq \p\left(\Vec{x}(t)|\hat{S}(t)=S_j\right),
\end{align}
where $a_{jk}$ is the time-invariant transition probability defined by the Markov chain, and $b_{j}(t)$ is the emission probability defined by the AAD algorithm. For the example Markov chain in Figure \ref{fig:HMM_model}, $a_{jk} = p_{switch}$ if $j\neq k$ and $1-(N-1)p_{switch}$ otherwise. The estimation of $ b_j (t)$ depends on the AAD algorithm, and is further explored in Section \ref{sec:AAD}.

Since a HMM assumes that $\p(\Vec{x}(t))$ only depends on $\hat{S}(t)$ and that $\p(\hat{S}(t)=S_j)$ only depends on $\hat{S(t-1)}$, we can write:
\begin{align}
    &\alpha_j(t) \triangleq\p\left(\hat{S}(t)=S_j,\Vec{x}_{0:t}\right)  \label{eq:ForwardAlgorithm}\\
        & = \sum_k \p\left(\hat{S}(t)=S_j, \Vec{x}_t,  \hat{S}(t-1) = S_k,\Vec{x}_{0:t-1}\right)\\
        & = b_j(t) \sum_k a_{kj} \alpha_k(t-1).
\end{align}
$\alpha_j(t)$ can thus be recursively computed for all speakers $S_j,\ j\in [1,N]$ and for each window $t$. At window $t=0$, we define $\alpha_j(0) = b_j(0)P(S(0)=S_j)=b_j(0)/N$, i.e., it is assumed that each speaker has an identical prior probability to be attended. The probability $\p(\hat{S}(t)=S_j|\Vec{x}_{0:t})$ that the listener attends to speaker $S_j$ in window $t$ is then: 
\begin{equation}
    \p\left(\hat{S}(t)=S_j|\Vec{x}_{0:t}\right) = \frac{\alpha_j(t)}{\sum_k \alpha_k(t)}.
    \label{eq:StateProbabilityForward}
\end{equation}

\subsection{Non-causal inference of the attended speaker}
\label{subsec:non-causalInference}
In non-causal inference, observations recorded after the window $t$ are also available. This allows us to calculate the probability $\p(\hat{S}(t) = S_j|\Vec{x}_{0:T})$ using all available observations $\Vec{x}_{0:T}$ using the forward-backward algorithm \cite{baum_maximization_1970}:
\begin{align}
    &\p(\hat{S}(t) = S_j|\Vec{x}_{0:T})\\
    &\quad= \frac{\p\left(\Vec{x}_{t+1:T}|\hat{S}(t)=S_j,\Vec{x}_{0:t}\right)\p\left(\hat{S}(t)=S_j,\Vec{x}_{0:t}\right)}{\p(\Vec{x}_{0:T})}\\
    &\quad= \frac{\p\left(\Vec{x}_{t+1:T}|\hat{S}(t)=S_j\right)\alpha_j(t)}{\sum_k \p(\Vec{x}_{t+1:T}|\hat{S}(t)=S_k)\alpha_k(t)}.
    \label{eq:forwardBackward_early}
\end{align}
$\alpha_j(t)$ can be recursively computed as in \eqref{eq:ForwardAlgorithm}. To simplify notations, we define:
\begin{equation}
    \beta_j(t) = \p(\Vec{x}_{t+1:T}|\hat{S}(t)=S_j),
\end{equation}
which can be understood as the probability to measure the observations $\Vec{x}_{t+1:T}$ given that the listener attends to speaker $S_j$ at window $t$. Similar to $\alpha_j(t)$, $\beta_j(t)$ can be computed recursively:

\begin{align}
    &\beta_j(t) \triangleq\p\left(\Vec{x}_{t+1:T}|\hat{S}(t)=S_j\right)\\
        & = \sum_k \beta_k(t+1)\p\left(\Vec{x}_{t+1}|\hat{S}(t+1)=S_k\right)a_{jk}\\
        & = \sum_k\beta_k(t+1)b_k(t+1)a_{jk}.
        \label{eq:backward}
\end{align}
The backward recursion starts at $\beta_j(T) = 1\, \forall j$. Substituting \eqref{eq:backward} in \eqref{eq:forwardBackward_early} then gives: 
\begin{equation}
    \p(\hat{S}(t) = S_j|\Vec{x}_{0:T}) = \frac{\alpha_j(t)\beta_j(t)}{\sum_k \alpha_k(t)\beta_k(t)}.
    \label{eq:forward-backward_Late}
\end{equation}

\subsection{Optimal sequence of attention states}
\label{subsec:Viterbi}
The most probable sequence of attention states can be computed with the Viterbi algorithm \cite{viterbi_error_1967}:
\begin{enumerate}
    \item Initialise $V_j(0) = 1/N$.
    \item Compute $V_j(t) = \max_k[V_k(t-1)a_{kj}b_j(t)],\ \forall j,t$.
    \item The most probable sequence of attention states is then given by the Viterbi path, which can be found by starting at the state $j^*_T = \arg \max_j V_j(T)$ and backtracking the path that was used to reach $V_j(T)$.
\end{enumerate}
Similar to the forward-backward algorithm, this model is non-causal. As we will show in Section \ref{sec:results}, it can predict the attended speaker with high accuracy and generally detects an attention switch quickly. However, due to its discrete nature, it is not possible to estimate how certain the algorithm is about its prediction of a specific attention state $\hat{S}(t)$. This is often undesired, which is why we will focus more on the causal forward and non-causal forward-backward algorithms in this paper. 

\subsection{Practical considerations}
Since the number of outcomes increases exponentially, some of the probability variables (such as $\alpha_j(t), \beta_j(t)$, and $V_j(t)$) quickly descend below machine accuracy as the sequence of observations becomes longer. This can lead to severe numerical instability, even in smaller problems.

Therefore, for a practical implementation, only the logarithm of the probability $\p_{log}(.)$ should be used. However, there is no straightforward method to compute $\log(\sum\p(.))$ based on $\p_{log}(.))$ without first computing the numerically unstable $\p(.)$. Instead, given a set of log probabilities $\Vec{P}_{log} = [P_{log,1} \dots P_{log,N}]$, the sum can be computed in a numerically stable way using the following method \cite{mcelreath_statistical_2018}:
\begin{align*}
    &\log\left(\sum_i P_{log,i}\right) =\\
    &\quad \max \Vec{P}_{log} + \log\left(\sum_i\exp\left(P_{log,i}-\max \Vec{P}_{log}\right)\right).
\end{align*}

The numerically stable forward-backward algorithm is specified in Algorithm \ref{alg:Forward-Backward}. 
\begin{algorithm*}
    \caption{The forward-backward algorithm \cite{baum_maximization_1970} to predict the attended speaker using a Hidden Markov Model in a numerically stable way. This algorithm is non-causal. For the causal algorithm from Section \ref{subsec:causalInference}, skip the backward pass and set all $\beta_{log,j}(t)=0$. The transition probabilities $a_{jk}$ represent the probability to switch from speaker $S_j$ to speaker $S_k$. The emission probabilities $b_j(t)$ represent the probability that speaker $S_j$ is attended in window $t$ and are generated by an AAD algorithm (see Section \ref{sec:AAD}.)}
    \label{alg:Forward-Backward}
    \KwIn{Transition probabilities $a_{jk}$, emission probabilities $b_j(t)$}
    \KwOut{$\p(\hat{S}(t)=S_j|\Vec{x}_{0:T}),\ \forall t \in [0,T]$}
    \vspace{0.5em}
    \textit{Initialize} $\forall j \in [1,N]$:\\
    $\alpha_{log,j}(0) \gets \log(b_j(0)/N)$\\
    $\beta_{log,j}(T) \gets 0$\\
    \vspace{0.5em}
    \textit{Complete the forward pass}:\\
    \For{$t = 1$ to $T$}{
        \For{$j=1$ to $N$}{
            \For{$k=1$ to $N$}{
                $\Vec{P}_{log,\alpha}(k) = \alpha_{log,k}(t-1) + \log(b_j(t)) + \log(a_{kj})$
            }
            $P_{\alpha,max} = \max \Vec{P}_{log,\alpha}$\\
            $\alpha_{log,j}(t) = P_{\alpha,max} + \log\left(\sum_k\exp(\Vec{P}_{log,\alpha}(k) - P_{\alpha,max}) \right)$
        }
    }
    \vspace{0.5em}
    \textit{Complete the backward pass}:\\
    \For{$t = T-1$ to $1$}{
        \For{$j=1$ to $N$}{
            \For{$k=1$ to $N$}{
                $\Vec{P}_{log,\beta}(k) = \beta_{log,k}(t+1) + \log(b_k(t+1)) + \log(a_{jk})$
            }
            $P_{\beta,max} = \max \Vec{P}_{log}$\\
            $\beta_{log,j}(t) = P_{\beta,max} + \log\left(\sum_k\exp(\Vec{P}_{log,\beta}(k) - P_{\beta,max}) \right)$
        }
    }
    \vspace{0.5em}

    \textit{Compute the probability that a speaker is attended $\p(\hat{S}(t)=S_j|\Vec{x}_{0:T})$}:\\
    \For{$t = 0$ to $T$}{
        \For{$j=1$ to $N$}{
            $\bm{\gamma}_{log}(j) = \alpha_{log,j}(t) + \beta_{log,j}(t)$\\
        }
        $\gamma_{max} = \max \bm{\gamma}_{log}$\\
        $\p_{log}(\hat{S}(t)=S_j|\Vec{x}_{0:T}) = \bm{\gamma}_{log}(j) - \gamma_{max}-\log\left(\sum_j\exp(\bm{\gamma}_{log}(j)-\gamma_{max})\right)$\\
        $\p(\hat{S}(t)=S_j|\Vec{x}_{0:T}) = \exp(\p_{log}(\hat{S}(t)=S_j|\Vec{x}_{0:T})$
    }
    \Return $\p(\hat{S}(t)=S_j|\Vec{x}_{0:T}),\ \forall t \in [0,T]$
    
\end{algorithm*}

%% AAD algorithms
\section{Auditory Attention Decoding}
\label{sec:AAD}
We consider an auditory attention decoding (AAD) model provides the scores $\Vec{x}(t) = [x_1(t) \dots x_D(t)]$, which are used as an input for the postprocessing Algorithm \ref{alg:Forward-Backward}. It is assumed that this score vector carries information about the attention state, i.e., the distribution of $\Vec{x}(t)$ changes with the state in the Markov chain. For instance, in AAD models that rely on envelope tracking \cite{OSullivan2014, DeCheveigne2018, Biesmans2017}, there is typically a single entry $x_s(t)$ per speaker (i.e., $D=N$) which refers to the correlation between the decoded envelope from the EEG and the actual speech envelope of speaker $s$. The entry $x_s(t)$ is expected to be high when speaker $s$ is attended and low otherwise. 

In the remaining of this paper, we will focus on such correlation-based score vectors generated by two popular envelope tracking AAD models: the least-squares (LS) algorithm first proposed in \cite{OSullivan2014} and refined in \cite{Biesmans2017}, and the canonical component analysis (CCA) algorithm \cite{DeCheveigne2018}. It is nonetheless important to stress that this is without loss of generality, as our framework is applicable to any type of score vector x(t), as long as a generative probabilistic model for the scores x(t) conditioned on the attention states is available\footnote{If such a model is not directly available, one can always train a generic generative model by fitting a Gaussian mixture model on score vectors of each state using the expectation maximization algorithm, or using the Baum-Welch algorithm if there are no attention labels available\cite{baum_maximization_1970}.}.

The least-squares algorithm computes a spatio-temporal transformation $\Vec{d}$ that attempts to reconstruct the envelope of the attended speech $y(t)$ as accurately as possible from the time-lagged EEG signal $\Vec{m}(t)$ \cite{OSullivan2014}:
\begin{equation}
    \Bar{\Vec{d}} = \argmin_\Vec{d} \E[(\Vec{d}^\top \Vec{m}(t) - y(t))^2],
    \label{eq:LS}
\end{equation}
where $\Vec{m}(t)$ is constructed as $[\Vec{m}_1^\top(t) \dots \Vec{m}_C^\top(t)]^\top$, with $\Vec{m}_c^\top(t) = [m_c(t) \dots m_c(t+L_m)]^\top$ a vector containing the EEG response at channel $c$ between $0$ and $L_m$ samples after $t$. Time lags are added to account for the delay in neural tracking as an auditory stimulus travels through the brain. Typically, an auditory stimulus can be detected for up to $\SI{250}{\milli\second}$ in the neural response \cite{OSullivan2014, Biesmans2017}, such that $L_m$ can be set to $0.25/f_s$ with $f_s$ the sampling rate. 

The CCA algorithm transforms both the EEG and the attended envelope, such that they are maximally correlated in a latent space \cite{DeCheveigne2018}:
\begin{equation}
    \Bar{\Vec{d}},\Bar{\Vec{e}} = \argmax_{\Vec{d},\Vec{e}} 
        \frac{
        \E[ \Vec{d}^\top\Vec{m}(t)\Vec{y}(t)^\top\Vec{e} ]}{
        \E [\|\Vec{d}^\top\Vec{m}(t)\|_2\|\Vec{e}^\top\Vec{y}(t)\|_2].=,
        }
    \label{eq:CCA}
\end{equation}
where $\Vec{m}(t)$ is defined as in \eqref{eq:LS}. $\Vec{y}(t) = [y(t-L_y) \dots y(t)]^\top$ is the time-lagged envelope of the attended speaker. Similar to $L_x$, $L_y$ is typically set at $\SI{250}{\milli\second}$. 

In general, multiple orthogonal decoder pairs $\Bar{\Vec{d}},\Bar{\Vec{e}}$ are computed for CCA. Each pair generates a single correlation per decision window. These correlation vectors are then typically transformed into 1-dimensional scores using a classification algorithm such as linear discriminant analysis (LDA) \cite{DeCheveigne2018}.

The LS and CCA models are both first trained on labelled training data. At test time, the correlation between the transformed EEG and the speech envelope of each speaker is calculated within a test window. The higher these correlations are, the more likely the speaker is to be attended. This process is repeated for each adjacent, non-overlapping\footnote{Overlap is to be avoided to ensure that the scores $\Vec{x}(t)$ and $\Vec{x}(t+1)$ are independent when conditioned on $S(t)$, which is a necessary assumption for obtaining \eqref{eq:ActualUnknownHMMProb} \cite{baum_maximization_1970}.} test window, generating a series of correlations $\Vec{x}(t) = [x_1(t) \dots x_S(t)]^\top$ for each speaker at each window $t$.

It now suffices to compute a sufficiently good estimation of the probability $b_j(t) \triangleq \p(\Vec{x}(t)|\hat{S}(t) = S_j)$ that speaker $j$ is attended given the correlations $\Vec{x}(t)$ at time $t$. This is done in a similar fashion as proposed in \cite{heintz_probabilistic_2024}. After the Fisher transformation $x' \leftarrow \atanh(x)$, Pearson correlations are approximately normally distributed \cite{fisher_frequency_1915}. This allows us to model the transformed correlations as two normal distributions:

\begin{align}
    \atanh(x_j) &\sim \mathcal{N}(\mu_{attended},\sigma^2)     & \text{if attended}\\
    \atanh(x_j) &\sim \mathcal{N}(\mu_{unattended},\sigma^2)   & \text{otherwise}.
\end{align}
This allows us to compute the likelihood $\p(x_j(t)|\hat{S}(t) = S_j)$ and $\p(x_j(t)|\hat{S}(t) \neq S_j)$ that $x_j(t)$ is observed if speaker $S_j$ is respectively attended and unattended. $\mu_{attended}$, $\mu_{unattended}$, and $\sigma$ can be estimated on training data. This can be done in a supervised fashion (e.g., based on labeled data from a different participant) or using an unsupervised estimator such as the techniques used in \cite{lopez-gordo_unsupervised_2025}.

We can then compute the emission probability $b_j(t) = \p(\Vec{x}(t) | \hat{S}(t) = S_j)$, as required in Algorithm \ref{alg:Forward-Backward}:
\begin{equation}
    b_j(t) = \p(x_j(t)|\hat{S}(t) = S_j) \prod_{k \neq j}\p(x_k(t)|\hat{S}(t) \neq S_k).
\end{equation}

Many more recent AAD algorithms rely on a classification algorithm to detect the attended speaker. In such cases, the output of the classifier can be directly used as $b_j(t)$ after applying a softmax function. However, the best way to translate these scores to the emission probability $b_j(t)$ depends heavily on the exact classification scheme. 

%% Experiments and data
\section{Experiments}
\label{sec:Experiments}
\subsection{Datasets}
The HMM algorithm is validated on two publicly available datasets. 

The first dataset is from \cite{rotaru_what_2024, rotaru_audiovisual_2024}, where a control for eye gaze was introduced. It contains 13 Flemish-speaking participants, who are instructed to attend to one of two competing speakers. The competing speakers are presented through insert phones in the left and right ears to the participant. An arrow on the screen indicated whether the participant had to listen to the left or right speaker. In each 10-minute trial, the participant switched attention to the other speaker after 5 minutes. This process is repeated in four trials. In two trials, the eye gaze of the participant was controlled such that it was uncorrelated with the location of the attended speaker. In one trial, the eye gaze was steered towards the location of the attended speaker. In one trial, the eye gaze was not controlled. We will not differentiate between these four settings, as it is mostly spatial AAD decoders that are affected by such gaze biases \cite{rotaru_what_2024}, whereas we use a linear stimulus reconstruction decoder (LS or CCA), which cannot use the gaze as a shortcut \cite{geirnaert_linear_2024}.  The 64-channel EEG is captured using a BioSemi ActiveTwo system. The original dataset also contains 4 EOG channels, but these are not used in the experiments in this paper. We will refer to this dataset as the '2 speaker' dataset.

In the second dataset, 20 participants must attend one of three simultaneous two-speaker conversations \cite{van_de_ryck_eeg-based_2025}. We will refer to this dataset as the '3 conversation' dataset. The conversations consisted of Flemish podcasts between two turn-taking speakers. The competing conversations were presented through loudspeakers and spatially separated by 27 degrees: one conversation in front, one 27 degrees left of the participant, and one 27 degrees right of the participant. There were six trials of ten minutes each in total. In three trials, the participants were instructed to sustain attention to a single conversation (once for each direction) for the entire duration of the trial. In the other three trials, the participants switched attention after five minutes. During the experiment, the  64-EEG was recorded with a BioSemi ActiveTwo system using a 10-20 layout. For more details, we refer to \cite{van_de_ryck_eeg-based_2025}.

For both datasets, we preprocessed the EEG and auditory stimuli according to the preprocessing framework proposed in \cite{Biesmans2017}. The auditory stimulus is split into 15 frequency bands using a gammatone filter bank with center frequencies between \SI{150}{\hertz} and \SI{4000}{\hertz}. The envelope is then extracted by computing $\sum_b |y_b(t)|^{0.6}$, with $y_b(t)$ the stimulus subbands obtained through the gammatone filter bank. This process is repeated for each auditory stimulus. The EEG and envelopes are then bandpass filtered between \SI{1}{\hertz} and \SI{9}{\hertz} and resampled to a sampling frequency of \SI{64}{\hertz}. 

We specifically define an attention switch as a listener switching between two speakers/conversations that occur at the same time. When two people speak intermittently in a conversation, the speech of these two speakers can be considered to be a single conversation, as proposed in \cite{van_de_ryck_eeg-based_2025}. We, therefore, do not consider switching between speakers within a single conversation to be an attention switch.

Although most experiments in this paper are performed using the original attention switches present after every 5 minutes in both datasets, artificial attention switches were created for the trials in the 3 conversation dataset without attention switches. This is done by permuting the order of the conversations every 5 minutes, such that the attended speech is each time at a different location. A similar methodology is used when we wish to test the HMM on data with more frequent attention switches. The HMM model does not perform significantly differently on artificial switches compared to original attention switches ($p = 0.87$, Wilcoxon signed rank).

\subsection{Evaluation criteria}
\label{subsec:EvalCriteria}
An HMM-based prediction is deemed correct if the speaker to whom the HMM assigns the highest probability (based on \eqref{eq:StateProbabilityForward} or \eqref{eq:forward-backward_Late} matches the true attended speaker. Using this definition, we consider two main metrics to evaluate the performance of the HMM model: the percentage of time the true attended speaker is correctly predicted by the HMM and the switch detection time; the average time between when a listener switches attention and when the newly attended speaker is correctly predicted by the HMM (i.e., when it is assigned the highest probability across all $N$ speakers). All predictions between the last switch detection and the next (ground-truth) attention switch are used to compute the accuracy in steady state. If no switch is detected before the next actual attention switch (which is rare), the switch detection time is defined as the time between the two actual ground truth switches and the entire block of data is used to compute the accuracy. This ensures that the accuracy of a model that never detects an attention switch remains low.

Since the non-causal model has access to future data, it can predict the switch before it happens in reality, which results in a negative switch detection time. Therefore, we will always report the average absolute value of the switch detection time, which can be viewed as the average error on the predicted timestamp of an attention switch.

The performance of the HMM model is influenced by various environmental and design parameters. These parameters are all investigated below. To keep all tests comparable, only one parameter is changed each time, while we use the baseline values specified in Table \ref{tab:baselineParam} for all other parameters. All tests are performed on both the causal and the non-causal (forward-backward) HMM models. These two models are suited for completely different applications (i.e., real-time attention tracking versus post-hoc attention tracking) and should not be directly compared. 

\begin{table}[t]
    \centering
    \begin{tabular}{c|c}
        Parameter & Base value \\
        \hline
        Dataset & 2 speakers \\
        Window length & \SI{1}{\second}\\
        Switch probability $p_{switch}$ & 0.001 * window length [s]\\
        Time between switches & \SI{5}{\minute}\\
        AAD algorithm & CCA \cite{DeCheveigne2018}
    \end{tabular}
    \caption{The default parameter values used in all experiments. In each experiment, only one parameter is changed, while all other parameters are kept constant.}
    \label{tab:baselineParam}
\end{table}

The AAD algorithms are trained and tested using a participant-specific leave-one-trial-out cross-validation strategy. The statistical parameters $\mu_{attended}$, $\mu_{unattended}$ and $\sigma$ are estimated using leave-one-participant-out cross-validation\footnote{We reiterate that there also exist unsupervised methods to estimate these variables, even on unlabelled data from the participant under test \cite{lopez-gordo_unsupervised_2025}. This is, however, beyond the scope of this paper.}. The HMM then predicts the attended speaker per window using all available AAD predictions within the same trial. 

\subsection{Causal and non-causal HMM}
In a first experiment, we will closely investigate the performance of the causal and non-causal HMM models using only baseline parameter values. For the non-causal HMM model, we investigate both the Forward-Backward algorithm and the Viterbi algorithm explained in Section \ref{sec:HMM}. We will show that the Viterbi algorithm has little benefit over the Forward-Backward algorithm for AAD. Therefore, we will uniquely focus on the Forward-Backward algorithm for non-causal HMM's in future experiments. 

The causal HMM is compared to three existing causal postprocessing methods for AAD: a simple Markov chain model where the probability that a speaker is attended increases with a fixed amount when the AAD algorithm predicts this speaker to be attended \cite{Geirnaert2020}, a state-space model that linearly scales the step size with the estimated probability that the speaker is attended \cite{heintz_probabilistic_2024} and a more complex state-space model where the decisions are smoothed using a Bayesian filter \cite{Miran2018}. 

\subsection{Influence of Window length}
AAD algorithms predict the attended speaker on a window-by-window basis. The length of these windows have a significant impact on the performance: when an AAD algorithm is used in isolation, short windows typically lead to a worse performance, but a faster detection of an attention switch \cite{Geirnaert2020,Geirnaert2020a}. In this experiment, we will study whether this trade-off is also present when the AAD algorithm is paired with a hidden Markov model. The window length is varied between \SI{0.25}{\second} and \SI{30}{\second}. The switching probability $p_{switch}$ between two decisions is proportionally changed such that $p_{switch} = 0.001\tau$, with $\tau$ the window length.

\subsection{Influence of AAD algorithms}
Naturally, the quality of the AAD algorithm is crucial for the HMM to work accurately. We will compare the LS and CCA algorithms to verify whether these two common stimulus reconstruction algorithms perform differently when combined with our proposed HMM post-processing algorithm. The LS algorithm uses a spatio-temporal filter with lags between \SI{0}{\milli\second} and \SI{250}{\milli\second} for the EEG. The CCA algorithm uses lags between \SI{0}{\milli\second} and \SI{250}{\milli\second} for the EEG and between \SI{-250}{\milli\second} and \SI{0}{\milli\second} for the stimulus. We use a total of 10 orthogonal decoder pairs, creating 10 different correlations per window. These are then classified with an LDA classifier, similar to \cite{DeCheveigne2018}. 

\subsection{Influence of switch probability}
There is a key trade-off in the hidden Markov model: the more stable it is, the less accurately it will detect an attention switch. This trade-off is controlled by the probability $p_{switch}$ that a listener switches attention to another speaker. Although $p_{switch}$ represents a probability within the HMM framework, it should be regarded as a tunable hyperparameter to control this trade-off, rather than an estimate of the true switching probability. If $p_{switch} \simeq 1/N$, the hidden Markov model does not impose any structure on the data: it will immediately switch to whichever speaker is most likely to be attended at time $t$ based solely on the data in window $t$. In contrast, as $p_{switch}$ approaches zero, the model is expected to only switch speakers when the AAD algorithm presents overwhelming evidence that a switch occurred over a prolonged period of time. This interaction is studied for both the causal (forward) and non-causal (forward-backward) algorithms in this experiment by varying $p_{switch}$ between $0.1$ and $0.0001$. 

\subsection{Influence of switching frequency}
The hidden Markov model assumes that the frequency of attention switches is relatively low. If the switching frequency rises above a certain threshold, the model would no longer be able to keep up and frequently miss attention switches entirely. This is especially relevant for the non-causal model, which also relies on future samples to predict the attended speaker. This mechanism is studied in this experiment by varying the switching frequency between \SI{20}{\second} and \SI{600}{\second}.  

\subsection{Influence of number of speakers/conversations}
We study in this experiment to what degree the HMM can generalise to listening environments with more than two speakers, such as the 3 conversations dataset. In such multi-speaker environments, each speaker is a different state, as illustrated in Figure \ref{fig:HMM_model}.

\subsection{Influence of the AAD accuracy}
In this experiment, we artificially control the accuracy of the CCA algorithm (without changing the window length) by altering the distance between the average attended and unattended correlations. More specifically, the new attended correlations are $x'_{attended} \leftarrow x_{attended} + \alpha (\mu_{attended}-\mu_{unattended})$, with $\alpha$ a factor between $-0.5$ and $5$.

This allows us to investigate how better AAD performances translate into better HMM performances. It can also give a lower bound for the minimal accuracy an AAD algorithm should achieve at a pre-defined window length to be sufficiently good for a specific application. 

%% Results
\section{Results and discussion}
\label{sec:results}
\subsection{Independence of AAD decisions}
The hidden Markov models assume that an AAD prediction only depends on the attention state of the listener and the observed EEG. Neighbouring samples should thus have no direct effect: $\p(\Vec{x}(t)|S(t),\Vec{x}(t-1),\Vec{x}(t-2),\dots) = \p(\Vec{x}(t)|S(t))$. Otherwise \eqref{eq:ActualUnknownHMMProb} (on which the entire HMM model is based) is no longer valid. This is approximately true for the predictions of both LS and CCA algorithms. Within a single attention state, the correlation of $x_{s_{attended}}(t)$ and $x_{s_{attended}}(t+1)$ is $0.1$ for both algorithms. The correlation between any $x_{s_{attended}}(t)$ and $x_{s_{attended}}(t+\tau)$, $\tau > 1$ is below $0.01$. 

However, AAD predictions from spatial algorithms such as CSP \cite{Geirnaert2020CSP} are typically much more correlated. The CSP classification scores for the attended speaker $\Vec{x}_{s_{attended}}(t)$ from two neighbouring windows have an average correlation of $0.88$\footnote{The correlation was computed on left-attended and right-attended speakers separately and then averaged. There was no discernible difference between the correlation of left-attended and right-attended speakers.}. This correlation steadily decreases as the distance between windows increases. However, the average correlation remains over $0.5$ when the windows are $\SI{20}{\second}$ apart. This large correlation between decisions is likely caused by the trial dependencies and eye gaze biases that majorly influence CSP \cite{rotaru_what_2024}. These dependencies would lead to major instabilities in the HMM, making spatial algorithms such as CSP less suited. 

\subsection{Causal and non-causal HMM}
Table \ref{tab:baselinePerformance} shows the average accuracy and average time between an actual and detected switch for the proposed hidden Markov models and the postprocessing algorithms proposed in \cite{Geirnaert2020,heintz_probabilistic_2024,Miran2018}. The non-causal forward-backward algorithm significantly outperforms the Viterbi algorithm. Although the Viterbi algorithm is equally accurate when no attention switches occur, the singular focus of the Viterbi algorithm on the most probable path makes the algorithm more likely to completely miss an attention switch.

When comparing causal postprocessing techniques, the proposed causal HMM model is both more accurate and detects attention switches faster than the Markov chain \cite{Geirnaert2020} and Bayesian model \cite{heintz_probabilistic_2024}. Although the state-space model detects switches faster using the hyperparameters proposed in \cite{Miran2018}, it achieves a much lower accuracy. To enable a more direct comparison, we increased the switching probability $p_{switch}$ in the causal HMM until we obtained a similar switch detection time of $\SI{7.9}{\second}$. With these settings, the HMM achieves an $80\%$ accuracy and $\SI{7.9}{\second}$ switch detection time and thus substantially outperforms the complex and computationally expensive state-space model. 

\begin{figure}[h]
    \begin{subfigure}{\columnwidth}
        \centering
        \includegraphics[width=\linewidth]{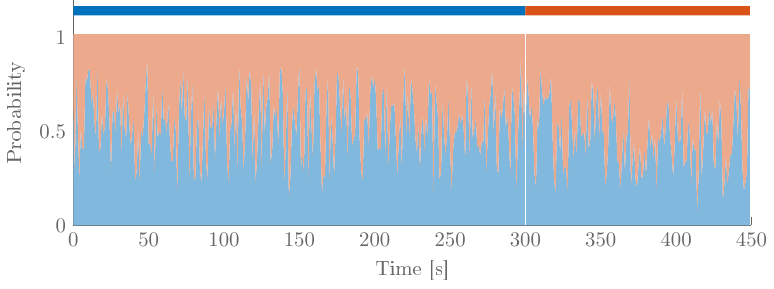}
        \caption{The raw CCA probabilities}
        \label{subfig:rawCCAProb}
    \end{subfigure}
    \par \bigskip
    \begin{subfigure}{\columnwidth}
        \centering
        \includegraphics[width=\linewidth]{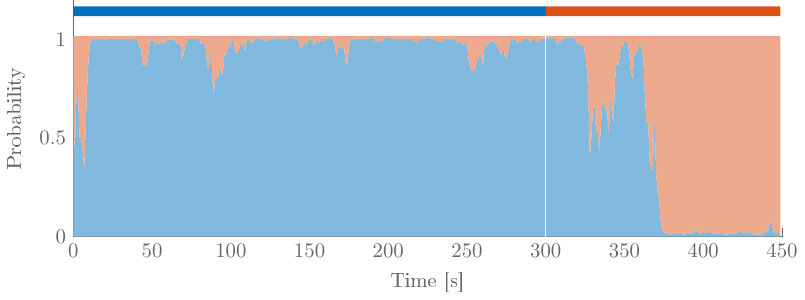}
        \caption{Causal HMM}
        \label{subfig:causalHMM}
    \end{subfigure}
    \par \bigskip
    \begin{subfigure}{\columnwidth}
        \centering
        \includegraphics[width=\linewidth]{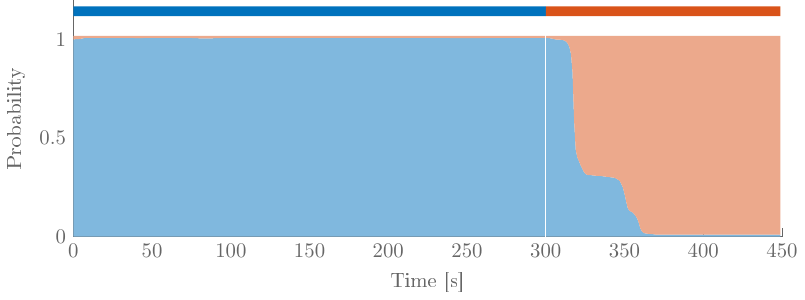}
        \caption{Forward-Backward HMM}
        \label{subfig:forwardBackwardHMM}
    \end{subfigure}
    \par \bigskip
    \begin{subfigure}{\columnwidth}
        \centering
        \includegraphics[width=\linewidth]{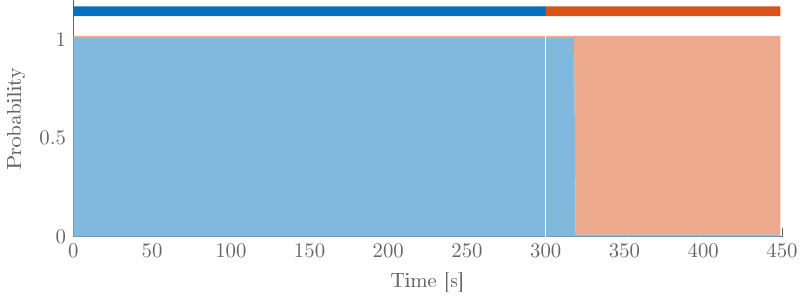}
        \caption{Viterbi HMM}
        \label{subfig:ViterbiHMM}
    \end{subfigure}
    \caption{A detail of the attention switch around \SI{300}{\second} of Participant 7 when the HMM models are combined with the CCA model. Participant 7 is an average-performing participant in the first dataset, and therefore a good representation of the broader population. The thin ribbon at the top represents the ground truth. The shaded area is directly proportional to the estimated likelihood that a specific speaker is attended.}
    \label{fig:AttentionSwitch}
\end{figure}

Figure \ref{fig:AttentionSwitch} demonstrates the behaviour of the three HMM models on an average-performing participant around the first switch in attention. The raw probabilities that are directly extracted from the CCA algorithm (Fig. \ref{subfig:rawCCAProb}) are clearly not suited as a decision variable as they continuously switch between both speakers (i.e., crossing the 0.5 threshold) even when there is no actual attention switch. The HMM-based predictions (Fig.\ref{fig:AttentionSwitch} \subref{subfig:causalHMM}-\subref{subfig:ViterbiHMM}) are much more reliable. Because the causal HMM model has no access to future samples, its estimated probabilities of the attended speaker are typically more jagged in steady-state. Furthermore, it is more susceptible to short bursts of inaccurate AAD predictions, as visible around $\SI{350}{\second}$.

The Viterbi algorithm does not estimate the probability that a certain speaker is attended. Instead, it only outputs the most likely sequence of attended speakers in a binary sequence. It is therefore unable to reflect uncertainty when necessary, e.g., in order to apply a non-binary relative gain to both speakers. This is an important limitation for AAD. 

\begin{table}[b]
    \centering
    \begin{subtable}[h]{\columnwidth}
        \centering
        \begin{tabular}{l|c c}
        Model type & Accuracy & Switch detection time\\
        \hline
            Forward-Backward HMM       & $97.2\% (\pm 3.4\%)$      & $\SI{17.0}{\second} (\pm \SI{12.1}{\second})$\\
            Viterbi HMM                & $96.5 \% (\pm 7.6\%)$     & $\SI{38.5}{\second} (\pm \SI{78.7}{\second})$
        \end{tabular}
        \caption{Non-Causal algorithms}
    \end{subtable}
    
    \begin{subtable}[h]{\columnwidth}
        \centering
        \begin{tabular}{l|c c}
        Model type & Accuracy & Switch detection time\\
        \hline
        Causal HMM   (proposed)           & $89.0\% (\pm 4.2\%)$      & $\SI{20.3}{\second} (\pm \SI{8.6}{\second})$\\
        Markov Chain \cite{Geirnaert2020} & $87.7\% \pm3.5\%)$ & $\SI{35.3}{\second} (\pm \SI{27.3}{\second})$\\
        Bayesian model \cite{heintz_probabilistic_2024} & $87.8\% \pm2.5\%)$ & $\SI{22.0}{\second} (\pm \SI{16.1}{\second})$\\
        State-Space model \cite{Miran2018}& $59.8 \% (\pm6.4\%$) & $\SI{7.7}{\second} (\pm\SI{7.7}{\second})$
    \end{tabular}  
    \caption{Causal algorithms}
    \end{subtable}
    \caption{The average accuracy and time required to switch between speakers of various causal and non-causal AAD postprocessing models. }
    \label{tab:baselinePerformance}
\end{table}

\subsection{Influence of Window length}
\begin{figure}[ht]
    \centering
    \begin{subfigure}{\columnwidth}
        \centering
        \includegraphics[width=\linewidth]{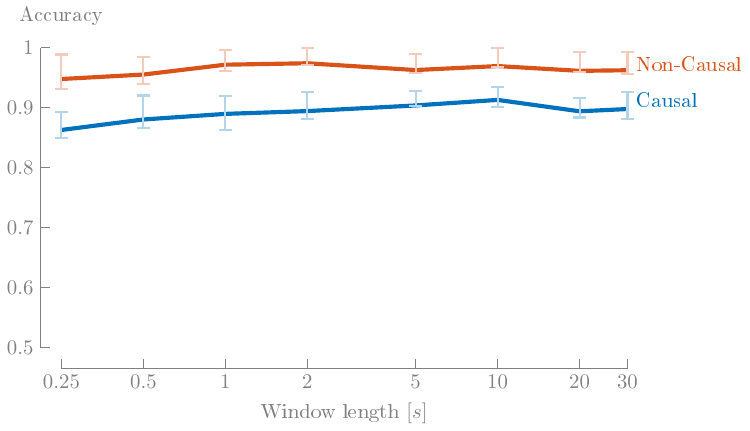}
        \caption{Accuracy}
    \end{subfigure}
    \begin{subfigure}{\columnwidth}
        \centering
        \includegraphics[width=\linewidth]{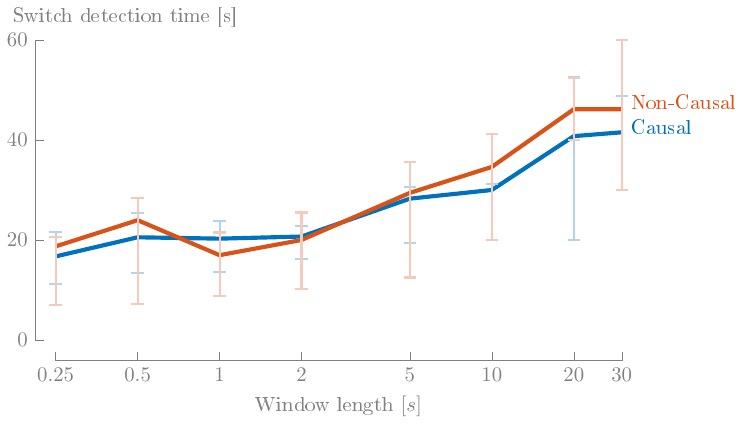}
        \caption{Time between actual and detected switch}
    \end{subfigure}
    \caption{Smaller window lengths enable HMMs to switch quickly between speakers without a major decline in accuracy.  The error bars represent the interquartile range.}
    \label{fig:windowLength}
\end{figure}

Although AAD algorithms are much less accurate on small window lengths, Figure \ref{fig:windowLength} demonstrates that this effect almost completely vanishes when they are combined with a HMM. On the contrary, the HMMs detect attention switches faster without a notable drop in accuracy when using shorter windows. However, there is a limit to this effect. Below \SI{1}{\second}, the AAD algorithms become too inaccurate to leverage the even faster decisions efficiently.

\subsection{Influence of AAD algorithms}
\label{subsec:AADalgorithm}
As shown in Table \ref{tab:LSversusCCA}, combining the HMM with a CCA model leads to significantly better results than with a the LS model. While the differences in accuracy are relatively small, the LS algorithm takes roughly 50\% longer to detect an attention switch ($p = 5e-4$, Wilcoxon signed rank).

This difference is quite striking, given the relatively small difference in performance between LS and CCA when used without HMM (the algorithms achieve respectively 57.7\% and 58.2\% accuracy on \SI{1}{\second} windows). This highlights how the HMM can rapidly improve with even a minor improvement in AAD accuracy, in particular when using short window lengths. This is further discussed in Section \ref{subsec:AADaccuracy}.  

\begin{table}[t]
    \centering
    \begin{tabular}{l|c c | c c}
    Model type & \multicolumn{2}{c}{Accuracy} & \multicolumn{2}{c}{Switch detection time}\\
    & LS & CCA & LS & CCA\\
    \hline
        Causal HMM              & $90.1\%$  & $89.0\%$ & $\SI{34.8}{\second}$ & $\SI{20.3}{\second}$\\
        Forward-Backward HMM     & $98.1\%$  & $97.2\%$ & $\SI{25.2}{\second}$ & $\SI{17.0}{\second}$\\
    \end{tabular}
    \caption{The accuracy and switch detection time of the causal and forward-backward HMM when paired with a least-squares (LS) and canonical component analysis (CCA) AAD algorithm.}
    \label{tab:LSversusCCA}
\end{table}

\subsection{Influence of switch probability}
\begin{figure}[ht]
    \centering
    \begin{subfigure}{\columnwidth}
        \centering
        \includegraphics[width=\linewidth]{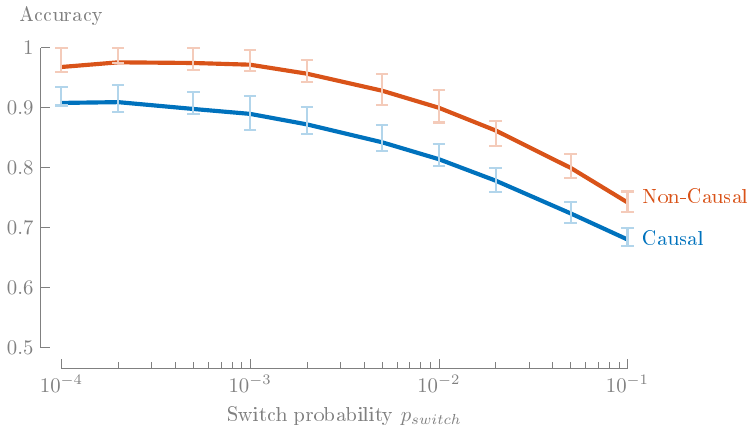}
        \caption{Accuracy}
    \end{subfigure}
    \begin{subfigure}{\columnwidth}
        \centering
        \includegraphics[width=\linewidth]{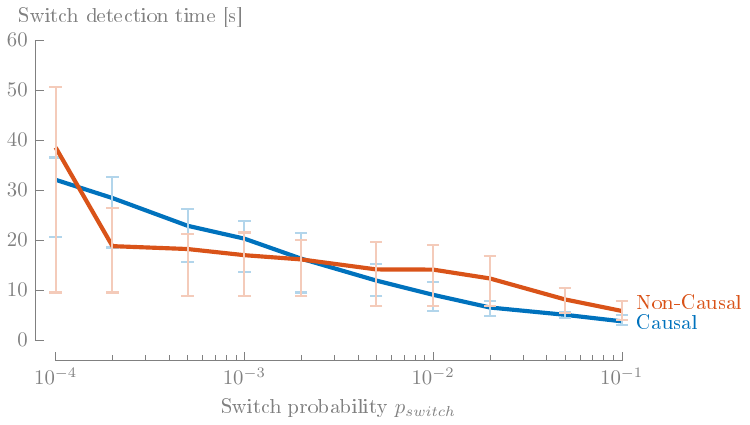}
        \caption{Time between actual and detected switch}
    \end{subfigure}
       \caption{The average accuracy and switch detection time become smaller as the switching probability increases. The error bars represent the interquartile range.}
    \label{fig:switchProb}
\end{figure}

Figure \ref{fig:switchProb} demonstrates that the HMMs decrease in accuracy, but detect attention switches faster, as the switching probability increases. As long as the switch probability remains sufficiently large, the non-causal HMM is less influenced by the switch probability than the causal HMM. However, at a specific threshold, the non-causal forward-backward model becomes increasingly mored likely to completely miss an attention switch. This leads to a sudden spike in switch detection time and drop in accuracy. 

\begin{figure}[h]
    \centering
    \includegraphics[width=\linewidth]{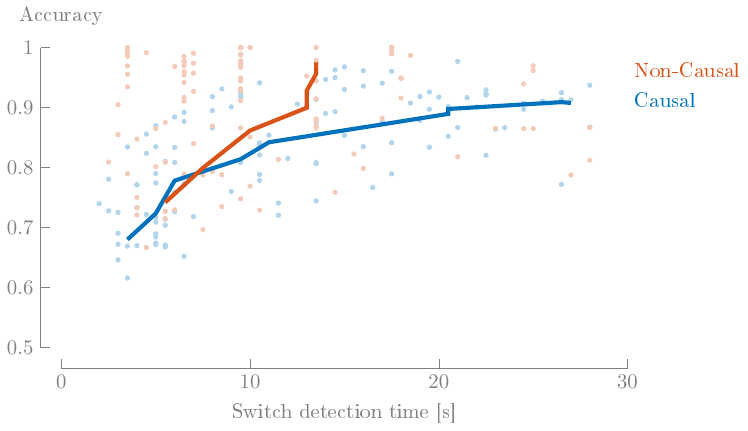}
    \caption{The trade-off between switching time and accuracy for hidden Markov models. The full line represents the average trade-off over all participants, while the coloured dots represent the average switching time and accuracy per participant and per tested value of $p_{switch}$.}
    \label{fig:accVSSpeedTrade-off}
\end{figure}

By computing the average time to switch speakers and average accuracy for each tested $p_{switch}$ value, it is also possible to directly investigate the trade-off between switching time and accuracy. This is shown in Figure \ref{fig:accVSSpeedTrade-off}. Both models clearly show the inverse relation between switching time and accuracy.

\subsection{Influence of switching frequency}
\begin{figure}[ht]
    \centering
    \begin{subfigure}{\columnwidth}
        \centering
        \includegraphics[width=\linewidth]{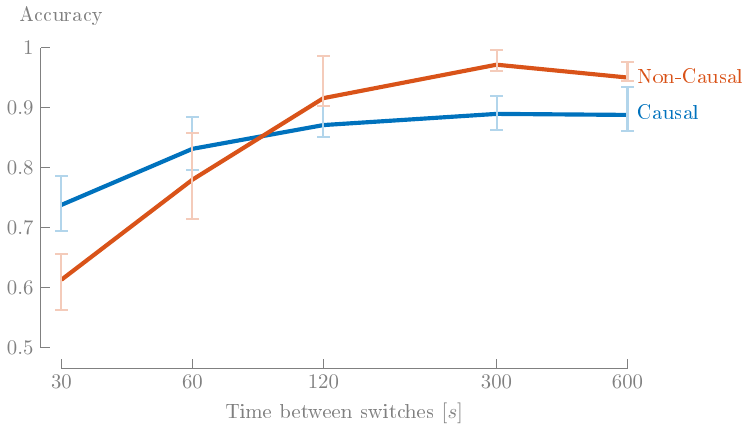}
        \caption{Accuracy}
    \end{subfigure}
    \begin{subfigure}{\columnwidth}
        \centering
        \includegraphics[width=\linewidth]{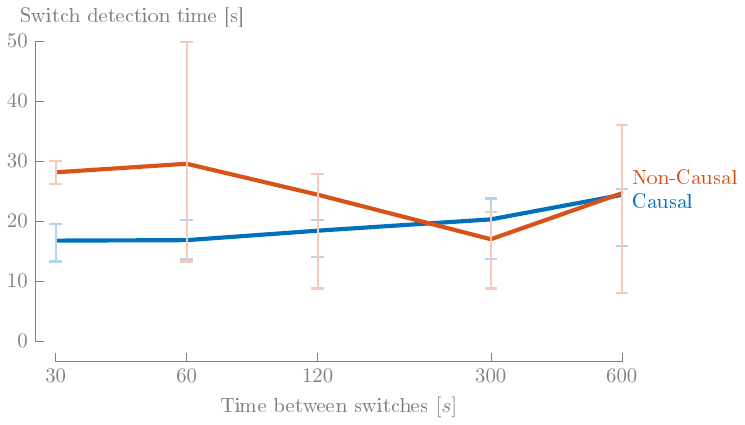}
        \caption{Time between actual and detected switch}
    \end{subfigure}
    \caption{Faster attention switches are harder to decode for hidden Markov models. Especially non-causal HMMs drop sharply in performance as attention switches become more frequent. The error bars represent the interquartile range.}
    \label{fig:Interval}
\end{figure}

The hidden Markov models drop significantly in accuracy as the time between switches decreases, as shown in Figure \ref{fig:Interval}. This is especially pronounced in the non-causal forward-backward algorithm. Consider an extreme scenario where the listener switched attention both briefly before and briefly after the investigated window. Since the non-causal algorithm has access to both past and future data, it detects that the unattended speaker is likely to be the attended speaker both shortly before and shortly after the investigated window. This will increase the estimated probability for the unattended speaker to be attended in the investigated window as well. When attention switches become too frequent, the non-causal algorithm will, therefore, miss more and more attention switches completely. This explains the dramatic drop in accuracy and increase in switch detection time. This effect can mostly be counteracted by increasing the switch probability $p_{switch}$ accordingly. However, this will make the HMM less accurate in steady-state. 

In general, the more frequently a listener switches attention, the harder it is for a HMM to accurately decode the true attention process. However, we reiterate that attention switches are specifically defined as a switch from one speaker or conversation to another speaker or conversation happening concurrently, which is not expected to happen frequently. The fast switches between turn-taking speakers partaking in a single conversation can be modeled as if the entire conversation is a single speaker.

\subsection{Influence of number of speakers}
Table \ref{tab:conversationPerformance} shows the accuracy of the 3 conversation dataset along with the 2 speaker dataset that was used until now.  As expected, the performance of the hidden Markov models drops when an additional competing speaker is added. This is mostly pronounced in the longer time required to switch between speakers. Remarkably, the accuracy barely decreases, even if we transitioned from a 2-class to a 3-class problem. However, these results should be interpreted with prudence. Apart from increasing the number of speakers, the dataset also uses a different experimental paradigm and interchanges single speakers with conversations, which could have an effect on the overall AAD performance. 

\begin{table}[b]
    \centering
    \begin{tabular}{l|c c | c c}
    Model type & \multicolumn{2}{c}{Accuracy} & \multicolumn{2}{c}{Switch detection time}\\
    Amount of speakers/conv. & 2 & 3 & 2 & 3\\
    \hline
        Causal HMM              & $89.0\%$  & $88.5\%$ & $\SI{20.3}{\second}$ & $\SI{52.7}{\second}$\\
        Forward-Backward HMM     & $97.2\%$  & $95.5\%$ & $\SI{17.2}{\second}$ & $\SI{37.2}{\second}$\\
    \end{tabular}
    \caption{The average accuracy and time required to switch between speakers on the dataset with 2 speakers and 3 conversations. Attention switches significantly slower in the dataset with three conversations compared to the dataset with two competing speakers. }
    \label{tab:conversationPerformance}
\end{table}

\subsection{Influence of the AAD accuracy}
\label{subsec:AADaccuracy}
\begin{figure}[ht]
    \centering
    \begin{subfigure}{\columnwidth}
        \centering
        \includegraphics[width=\linewidth]{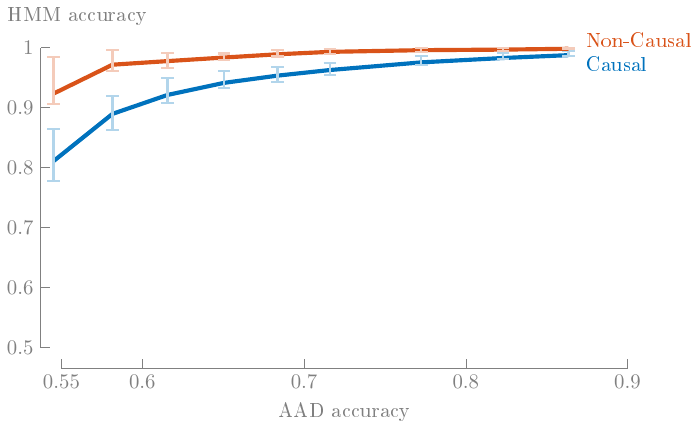}
        \caption{Accuracy}
    \end{subfigure}
    \begin{subfigure}{\columnwidth}
        \centering
        \includegraphics[width=\linewidth]{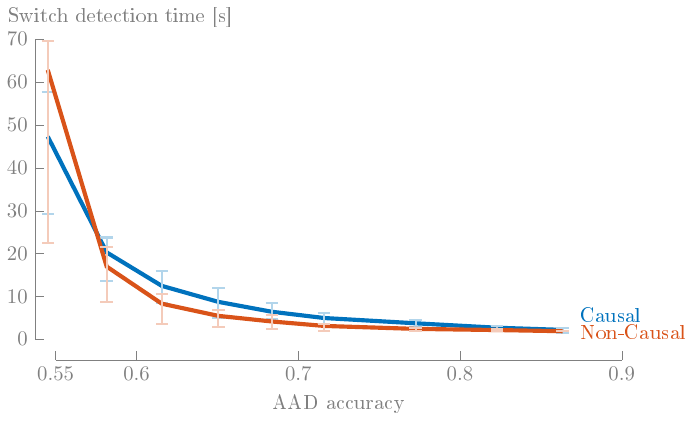}
        \caption{Time between actual and detected switch}
    \end{subfigure}
    \caption{A small improvement in AAD accuracy leads to a major improvement in HMM performance. The second-lowest AAD accuracy represents true performance. All other samples are artificially generated to model the potential improvement enabled by better AAD algorithms or EEG sensors.}
    \label{fig:aadAccuracy}
\end{figure}

As already hinted in Section \ref{subsec:AADalgorithm}, even a small improvement in AAD accuracy leads to a major improvement in both the accuracy and the switch detection time of the HMM. This is shown in Figure \ref{fig:aadAccuracy}. An AAD accuracy of $60\%-65\%$ on \SI{1}{\second} windows would suffice for most practical applications when paired with a hidden Markov model. 

There are already various AAD algorithms that claim to achieve or surpass such performances \cite{Geirnaert2020CSP, Vandecapelle2020,su_stanet_2022}. However, these algorithms often fail to generalise well on new, unseen data or on data where the eye gaze is uncorrelated to the direction of attention, as they tend to overfit on trial- or gaze-related biases in the training dataset \cite{rotaru_what_2024, Puffay2023,ivucic_impact_2024}. Furthermore, the AAD predictions of these algorithms are often heavily correlated (due to the aforementioned biases), which would significantly decrease the beneficial effect of an HMM. To our knowledge, no AAD algorithm exists yet that was shown to outperform CCA on a gaze-controlled dataset such as that of \cite{rotaru_what_2024, rotaru_audiovisual_2024} and satisfies all requirements explained above. 

%% Conclusion
\section{Conclusion}
\label{sec:Conclusion}
An attention process is inherently structured: while a person could switch attention at any moment, they are, on average, much more likely to keep attending the same person. Hidden Markov models (HMMs) are an intuitive and computationally efficient way to utilise this structure. They can easily be combined with existing AAD algorithms, as long as the AAD score vectors $\Vec{x}(t)$ only depend on the attention state. 
The combination of a HMM and CCA was shown to be especially powerful, enabling average decoding accuracies above 90\% while detecting any attention switch within \SI{20}{\second} of the actual switch. While the model performs best when it has access to both previous and future data, it still outperforms state-of-the-art postprocessing models when only past data are available. This paves the way toward fast and accurate attention decoding, both offline and in real-time. 

\bibliographystyle{IEEEtran}
\balance
\bibliography{references}
\end{document}